\documentclass[aps,prd,amsmath,amssymb,superscriptaddress,twocolumn,nofootinbib,showpacs,preprintnumbers,notitlepage]{revtex4-1}
\usepackage[usenames,dvipsnames,svgnames,table]{xcolor}

\usepackage{subfigure,lineno,wrapfig}
\usepackage{graphicx}% Include figure files
\usepackage{dcolumn}% Align table columns on decimal point
\usepackage{bm}% bold math
\usepackage{hyperref}% add hypertext capabilities
\hypersetup{ 
    pdfnewwindow=true,      % links in new window
    colorlinks=true,       % false: boxed links; true: colored links
    allcolors=[RGB]{31 119 180}
} 
\usepackage[capitalise]{cleveref}
\usepackage{dsfont}
\usepackage{placeins}
\usepackage{todonotes}
\usepackage[utf8]{inputenc}
\usepackage{gensymb}
\usepackage{multirow}
\usepackage{amsmath}
\usepackage{amsfonts}
\usepackage{amssymb}
\usepackage{enumitem,amssymb}
\usepackage{array}   % for \newcolumntype macro
\newcolumntype{L}{>{$}l<{$}} % math-mode version of "l" column type
\newcolumntype{R}{>{$}r<{$}}
\newcolumntype{C}{>{$}c<{$}}
\usepackage{xspace}
\usepackage{braket}
\usepackage{units}
\usepackage{slashed}
\usepackage[normalem]{ulem}
\usepackage{tabularx}
\usepackage{nameref}
\usepackage{orcidlink}

\usepackage{xpatch}% http://ctan.org/pkg/xpatch
\makeatletter
\xpatchcmd{\@ssect@ltx}{\@xsect}{\protected@edef\@currentlabelname{#8}\@xsect}{}{}% Patch \<section>*
\xpatchcmd{\@sect@ltx}{\@xsect}{\protected@edef\@currentlabelname{#8}\@xsect}{}{}% Patch \<section>
\makeatother
\usepackage{hyperref}% http://ctan.org/pkg/hyperref

\let\Re\relax
\let\Im\relax
\DeclareMathOperator{\Re}{Re}
\DeclareMathOperator{\Im}{Im}

\newcommand{\ie}{{\it i.e.}\xspace}

\newcommand{\mevnospace}{\ensuremath{{\mathrm{\,Me\kern -0.1em V}}}}
\newcommand{\gevnospace}{\ensuremath{{\mathrm{\,Ge\kern -0.1em V}}}}
\newcommand{\tevnospace}{\ensuremath{{\mathrm{\,Te\kern -0.1em V}}}}
\newcommand{\mev}{\mevnospace\xspace}
\newcommand{\gev}{\gevnospace\xspace}

\usepackage{pifont}

\newcommand{\catania}{INFN Sezione di Catania, I-95123 Catania, Italy}
\newcommand{\ceem}{Center for  Exploration  of  Energy  and  Matter, Indiana  University, Bloomington,  IN  47403,  USA}
\newcommand{\gwu}{Department of Physics, The George Washington University, Washington, DC 20052, USA}
\newcommand{\hiskp}{Helmholtz-Institut f\"{u}r Strahlen- und Kernphysik (Theorie) and Bethe Center for Theoretical Physics, Universit\"{a}t Bonn, D-53115 Bonn, Germany}
\newcommand{\indiana}{Department of Physics, Indiana  University, Bloomington,  IN  47405,  USA}
\newcommand{\jlab}{Theory Center, Thomas  Jefferson  National  Accelerator  Facility, Newport  News,  VA  23606,  USA}
\newcommand{\lbnl}{Nuclear Science Division, Lawrence Berkeley National Laboratory, Berkeley, CA 94720, USA}
\newcommand{\messina}{Dipartimento di Scienze Matematiche e Informatiche, Scienze Fisiche e Scienze della Terra, Universit\`a degli Studi di Messina, I-98166 Messina, Italy}
\newcommand{\odu}{Department of Physics, Old Dominion University, Norfolk, Virginia 23529, USA}
\newcommand{\ub}{Departament de F\'isica Qu\`antica i Astrof\'isica and Institut de Ci\`encies del Cosmos, Universitat de Barcelona, Barcelona E-08028, Spain}
\newcommand{\ucb}{Department of Physics, University of California, Berkeley, CA 94720, USA}
\newcommand{\uned}{Departamento de F\'isica Interdisciplinar, Universidad Nacional de Educaci\'on a Distancia (UNED), Madrid E-28040, Spain}

\begin{document}
\preprint{JLAB-THY-24-4058}
\title{Nonperturbative aspects of the electromagnetic pion form factor at high energies}

\author{K.~\surname{Quirion}\orcidlink{0000-0003-3383-7774}}
\email{kquirion@iu.edu}
\affiliation{\indiana}
\affiliation{\ceem}
\author{C.~\surname{Fern\'andez-Ram\'irez}\orcidlink{0000-0001-8979-5660}}
\email{cefera@ccia.uned.es}
\affiliation{\uned}
\author{V.~\surname{Mathieu}\orcidlink{0000-0003-4955-3311}}
\affiliation{\ub}
\author{G.~\surname{Monta\~na}\orcidlink{0000-0001-8093-6682}}
\affiliation{\jlab}
\author{R.~J.~\surname{Perry}\orcidlink{0000-0002-2954-5050}}
\affiliation{\ub}
\author{A.~\surname{Pilloni}\orcidlink{0000-0003-4257-0928}}
\affiliation{\messina}
\affiliation{\catania}
\author{A.~\surname{Rodas}\orcidlink{0000-0003-2702-5286}}
\affiliation{\jlab}
\affiliation{\odu}
\author{V.~\surname{Shastry}\orcidlink{0000-0003-1296-8468}}
\affiliation{\indiana}
\affiliation{\ceem}
\author{W.~A.~\surname{Smith}\orcidlink{0009-0001-3244-6889}}
\affiliation{\messina}
\affiliation{\gwu}
\affiliation{\ucb}
\affiliation{\lbnl}
\author{A.~P.~\surname{Szczepaniak}\orcidlink{0000-0002-4156-5492}}
\affiliation{\indiana}
\affiliation{\ceem}
\affiliation{\jlab}
\author{D.~\surname{Winney}\orcidlink{0000-0002-8076-243X}}
\affiliation{\hiskp}

\collaboration{Joint Physics Analysis Center}

\begin{abstract}
The structure of hadronic form factors at high energies and their deviations from perturbative quantum  chromodynamics provide insight on nonperturbative dynamics.  Using an approach that is consistent with dispersion relations, we construct a model that simultaneously accounts for the pion wave function, gluonic exchanges, and quark Reggeization. In particular, we find that quark Reggeization can be investigated at high energies by studying scaling violation of the form factor.
\end{abstract}
\maketitle

\section{Introduction}\label{sec:intro}
Form factors encode information about hadron composition. For example, form factors at low energies have been used to discriminate between constituent quark and mesonic degrees of freedom~\cite{Perdrisat:2006hj,Roberts:2021nhw,Barabanov:2020jvn}, while at large momentum transfer they can be used to investigate the short distance interactions between partons inside a hadron. The pion electromagnetic form factor plays a special role because of the simplicity of the theoretical prediction~\cite{Li:1992nu,OConnell:1995nse,Stefanis:1998dg,Bakulev:2004cu,Gorchtein:2011vf,Perry:2018kok}.  At high energies, partons are asymptotically free.  Therefore, when the photon interacts with one of the partons of a hadron, it is possible for the large momentum transfer to be shared internally between a few constituents via hard QCD interactions in such a way that the soft wave functions are shielded from large momentum flows ~\cite{Brodsky:1973kr,Duncan:1979hi,Efremov:1979qk,Lepage:1980fj}. Hence, the contributions from hard and soft processes factorize, and as $Q^2 \to \infty $ one can rely on perturbative QCD (pQCD) to calculate the spacelike charged pion electromagnetic form factor~\cite{Farrar:1979aw,Efremov:1978rn,Lepage:1979zb}
\begin{align}
 F(Q^2) \to 16 \pi \frac{\alpha_s(Q^2)}{Q^2}  f_\pi^2 \, .\label{eq:pqcd}
\end{align}
Here $f_\pi \approx 93 \mev $ is the pion decay constant, $\alpha_s(Q^2)$ is the running QCD coupling, and $Q^2$ is the photon virtuality. The normalization of the form factor is determined by soft processes, \ie it relates to the pion wave function, which asymptotically approaches a universal form~\cite{Lepage:1979zb}.The perturbative result in the spacelike region can be analytically continued to the timelike region \mbox{$Q^2=-s$}, where $s$ is the energy squared of the final $\pi^+\pi^-$ system in its own center-of-mass reference system.\footnote{Throughout the paper, we will briefly refer to the dependence on $s$ as `energy dependence'. } Several measurements of the pion  and other pseudoscalar meson  form factors  have  been performed in the past~\cite{Brown:1973wr,Bebek:1974ww,JeffersonLabFpi:2000nlc,Aulchenko:2006dxz,KLOE:2004lnj,Aulchenko:2006dxz,Achasov:2006vp,JeffersonLabFpi-2:2006ysh,JeffersonLabFpi:2007vir,Horn:2007ug,CMD-2:2006gxt,BaBar:2009wpw,BaBar:2012bdw,BESIII:2015equ,KLOE:2010qei,Seth:2012nn}, with more experiments scheduled for the future~\cite{AbdulKhalek:2021gbh,Accardi:2023chb}. It was argued in Refs.~\cite{Isgur:1984jm,Isgur:1988iw,Belyaev:1995ya} that for $s\lesssim 10 \gev^2$ the soft contributions dominate, so the asymptotic behavior of~\cref{eq:pqcd} may not yet have been reached~\cite{Gousset:1994yh,Simula:2023ujs}. Indeed, the analysis of~\cite{Seth:2012nn} from \mbox{CLEO-c} finds two data points at $s=16.1\gev^2$ and $18.3\gev^2$ that are approximately a factor of two above the pQCD prediction. In light of this discrepancy between the asymptotic pQCD prediction and experimental data in the intermediate energy region currently accessible, it is interesting to study models that transition between soft and hard processes, which may also be more applicable to energy regions relevant to current experiments. Our goal is to explore what deformations of the building blocks of the perturbative calculation, that encode unknown nonperturbative dynamics, are compatible with the pQCD asymptotics. 
Specifically, we calculate the timelike form factor with a dispersive approach that allows us to use building blocks having different microscopic dynamics, and select the ones that lead to a result compatible with the correct asymptotic behavior. Among the various soft effects, we consider a Reggeized quark exchange, with the relative trajectory built self-consistently.  For simplicity, we treat all particles as scalars. Consequently, our calculations should not be interpreted as predictions for the  behavior of the pion form factor, but rather as an exploration of the soft mechanisms that one might observe for physical quarks.

The rest of the paper is organized as follows. \cref{sec:dispersive} introduces the dispersive approach. In~\cref{sec:PerturbativeModels} we present several microscopic mechanisms and their impact on the asymptotic behavior of the form factor. \cref{sec:ReggeizedModel} presents the Reggeized model. Finally, \cref{sec:conclusions} summarizes the results and conclusions.

\section{Dispersive approach}\label{sec:dispersive}

\begin{figure}
\includegraphics[width=0.35\textwidth]{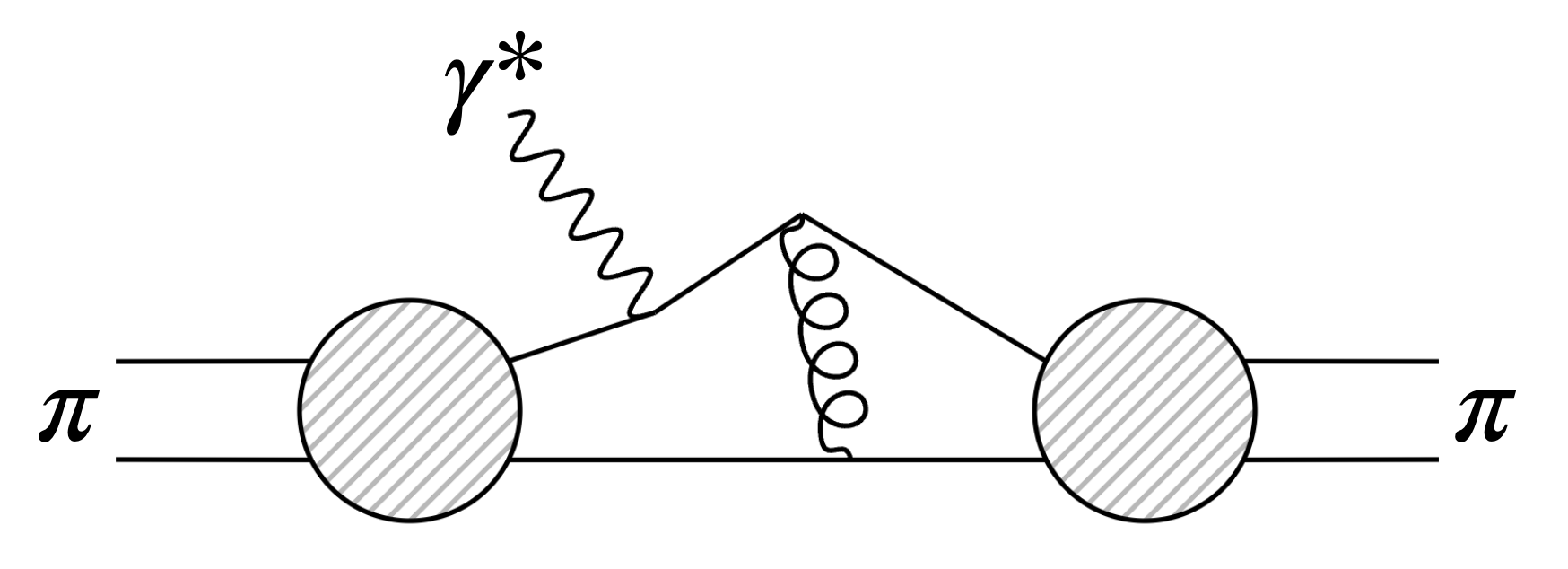}
\caption{ Spacelike electromagnetic pion form factor: pQCD mechanism requires a hard gluon exchange between the pion constituents~\cite{Lepage:1979zb}.}
\label{fig:spacelike}
\end{figure}  

\begin{figure*}
\includegraphics[width=0.8\textwidth]{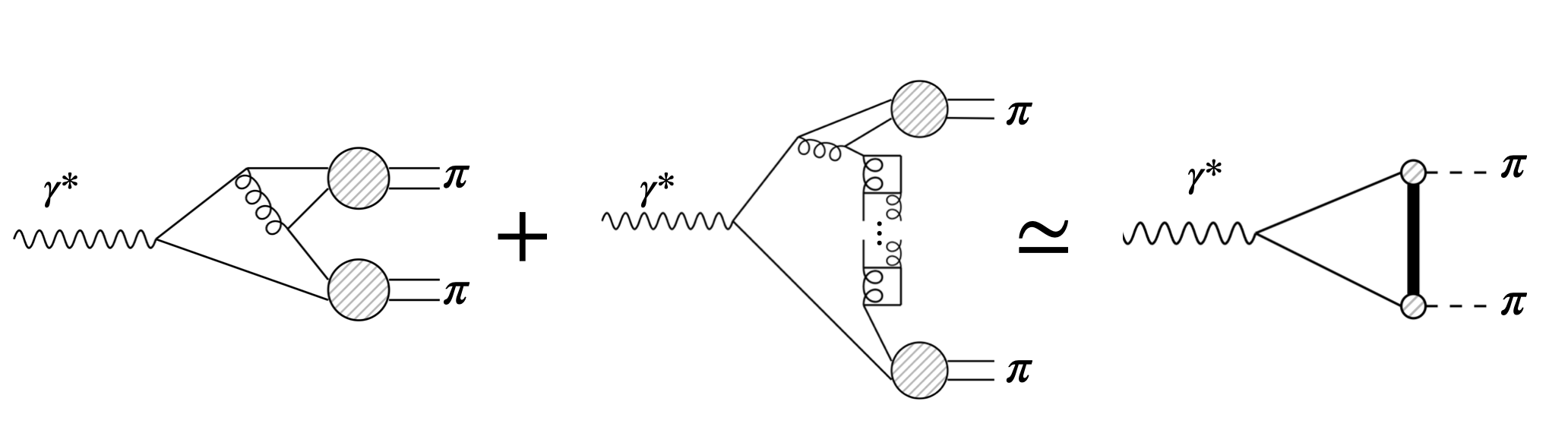}
\caption{Timelike electromagnetic pion form factor: the hard gluon required to create the additional $q\bar q$ pair can stream a quark-gluon ladder. Such Reggeized quark can reabsorb the unknown soft physics from the two-pion wave functions.}
\label{fig:timelike}
\end{figure*}  

We consider the electromagnetic form factor $F(s)$. In the spacelike region, the physics picture can be obtained from light-cone time ordered perturbation theory, as represented in \cref{fig:spacelike}: the virtual photon strikes one of the quarks from the incoming pion, which carries soft momentum according to the pion wave function. If no hard gluon is emitted, the probability of having one soft and one hard quark in the outgoing pion is suppressed by the wave function. The extra gluon allows the two constituents to rearrange their momenta, preventing the suppression. The gluon propagator provides the \mbox{$\alpha_s(Q^2)/Q^2$} in~\cref{eq:pqcd}~\cite{Lepage:1980fj}.
The form factor is thus proportional to the overlap of the incoming and outgoing pion wave functions. On the other hand, in the timelike region the picture is different, as represented in~\cref{fig:timelike}: a $q\bar q$ pair with high relative momentum is emitted by the photon. To convert into pions, we need at least another $q\bar q$ pair emitted by a hard gluon. In this frame, the form factor is incalculable, as the joint wave function of the four constituents inside the two pions is completely unknown. However, we know that in $q \bar q\to gg$ scattering at high energies, the quark exchanged in the $t$-channel can be dressed with gluons, filling the rapidity gap between the initial quarks. Higher order diagrams (ladders) carry large logarithms and need to be resummed, which leads to quark Reggeization~\cite{Gell-Mann:1964aya,Gell-Mann:1964ncb,Gribov:2009zz,Fadin:1977jr,Strikman:2007nz,Gorchtein:2011vf}. So, instead of focusing on unknown multibody wave functions, we can infer the energy dependence from scattering theory and Regge phenomenology.

\begin{figure}
\includegraphics[width=0.48\textwidth]{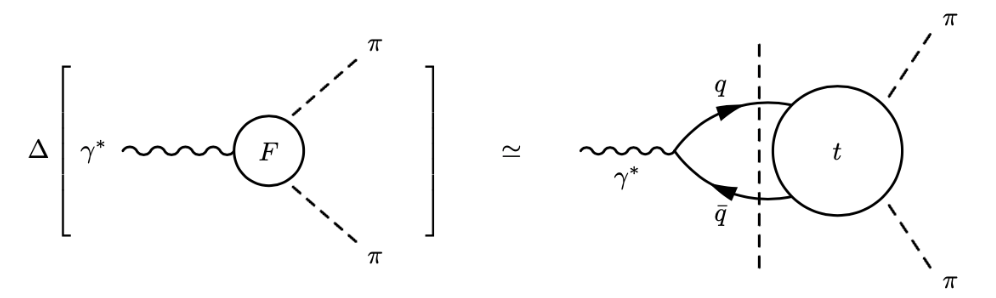}
\caption{Discontinuity  of the pion electromagnetic form factor $\Delta F(s)$ across the unitarity cut. The \mbox{$\gamma^* \to q \bar{q}$} amplitude is assumed to be pointlike and all the depicted particles are on shell.}
\label{fig:discontinuity}
\end{figure}

To keep the calculations simple, we treat all particles as scalars. We thus do not focus on the precise normalization, which would be different for physical quarks. We also ignore the dependence on $\alpha_s(s)$, which carries additional logarithmic dependence. For our purposes, we thus compare with the pQCD asymptotic result being \mbox{$F(s) \sim 1/s$}.
The form factor can be written in terms of its discontinuity across the unitary cut~\cite{Gorchtein:2011vf},  \mbox{$\Delta F(s)= \lim_{\epsilon \to 0}  \left[ F(s+ i \epsilon) - F(s - i\epsilon) \right]/2i$} using a dispersion relation
\begin{align}
 F(s) = \, \frac{1}{ \pi } \int_{4m_\pi^2}^\infty \, ds' \, \frac{\Delta  F(s')}{s' - s} \, . \label{eq:dicFormFactorGeneral}
\end{align}
In the high energy regime, as illustrated in~\cref{fig:discontinuity}, it can be approximated by
\begin{align}
F(s)  \simeq \, \frac{1}{ \pi } \int_{s_t}^\infty \, ds' \,  \frac{ \,  t^*_{q \bar{q} \to \gamma^*} \, \rho_{q\bar{q}}\, t_{q \bar{q} \to \pi \pi}}{s' - s} \, ,
\label{eq:dicFormFactor}
\end{align}
where $s_t$ is the threshold, and $t_{\gamma^*\to q\bar{q}}$ and $t_{q\bar{q}\to\pi\pi}$ are transition amplitudes from a virtual photon to a $q\bar q$ pair, and from the $q\bar q$ pair to two pions, respectively.\footnote{The same \cref{eq:dicFormFactor} would be obtained from Feynman diagrams with effective $qq\pi$ vertices using Cutkosky rules.} We note that the actual value of  $s_t$ is irrelevant, given that we are interested in the high energy behavior. We use the notation $\rho_{q\bar{q}}$ to represent the phase space factor for the production of the quark-antiquark pair. We are interested in understanding the role of various mechanisms contributing to $\Delta F(s)$ at high energies. 

Asymptotically  the two-body phase space  \mbox{$\rho_{q\bar q}(s\to\infty)  \to 1$}, thus at some large \mbox{$s \gg s_t$} the discontinuity originates from a $q\bar q$ intermediate state and \mbox{$\Delta F(s) \simeq  t^*_{q \bar{q}  \to \gamma^*} \, t_{q \bar{q} \to \pi \pi}$}. To be compatible with the pQCD asymptotics, we keep the photon vertex pointlike, as finite size modifications would typically lead to suppressions stronger than $1/s$. We thus focus on the  hadronization dynamics contained in the \mbox{$t_{q \bar q \to \pi\pi}$} amplitude.

To obtain the $s\to \infty$ behavior, the denominator of~\cref{eq:dicFormFactorGeneral} can be expanded, resulting in a sum of moments of the discontinuity of the form factor
\begin{align}
F(s) &=  -\frac{1}{ \pi } \int^s_{s_t} ds' \frac{\Delta  F(s')}{s} \left[ 1 + \frac{s'}{s} + \left( \frac{s'}{s}\right)^2 + \dots\right]  \nonumber \\
&+  \frac{1}{\pi } \int_s^\infty ds' 
\frac{\Delta F(s')}{s'} \left[ 1 + \frac{s}{s'} + \left( \frac{s}{s'}\right)^2 + \cdots\right].
\label{eq:disp_exp} 
\end{align}

\begin{figure}
\includegraphics[width=0.3\textwidth]{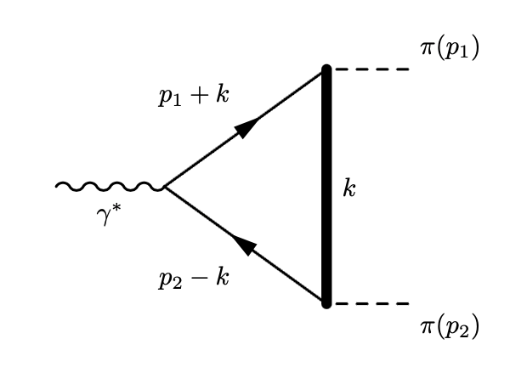}
\caption{Model for the electromagnetic pion form factor as \mbox{$\gamma^* \to q\bar{q} \to \pi\pi$}. The $q\bar{q} \to \pi\pi$ transition is mediated by the exchange $e(k)$, which could be either a quark or a Reggeon.} \label{fig:modelpert}
\end{figure}

The convergence of~\cref{eq:dicFormFactorGeneral} implies that the integral over $s'>s$ becomes  negligible as $s\to \infty$, and we can keep only the sum of moments of the first term.  Hence, if we were interested in an amplitude that behaves according to $s^{-n}$ at large $s$,  we should require that the first $n-1$ moments of the discontinuity of the form factor vanish, \ie
\begin{align}
\int^\Lambda_{s_t} ds' \, (s')^i \, \Delta F(s') = 0, \quad i = 0,1,\cdots n-1\, , \label{eq:zero_moments}
\end{align}
where $\Lambda \simeq s$. Therefore, a positive $\Delta F(s)$ leads to a $s^{-1}$ falloff (modulo logarithms) of the form factor. A falloff faster than $s^{-1}$ implies the existence of intermediate states $X$ other than $X=q\bar q$ which contribute destructively  to the discontinuity of $F(s)$. This is analogous to the cancellation found in Regge cut  models~\cite{Amati:1962ey}. 

As said in the introduction, the goal of this paper is to assess the impact of several nonperturbative mechanisms on the asymptotic behavior and, in particular, to understand which ones are compatible with the pQCD predictions. In what follows, we discuss the role of each individual mechanism.

\section{Individual mechanisms and their impact on the asymptotic behavior}\label{sec:PerturbativeModels}

\subsection{Pointlike}\label{sec:pert}
To establish a point of reference for the comparison of different microscopic contributions, we first consider a model, as depicted in~\cref{fig:modelpert}, where the $\gamma \to q\bar q$ interaction is pointlike, the exchanged particle $e$ with momentum $k$ is not Reggeized,  and there are no soft pion wave functions. This makes all the interactions as hard as possible. While this is not a realistic representation of the form factor, it provides an upper limit for the behavior of the physical form factor at high energies. In~\cref{fig:modelpert} the propagators are \mbox{$G(p,m) = (p^2 - m^2)^{-1}$} and the transition amplitude $q\bar{q} \to \pi \pi$ is mediated by a (scalar) quark exchange. Hence, disregarding the normalization, we can write the pointlike form factor $F_0(s)$ as
\begin{align}
F_0(s) = \frac{1}{i} \int \frac{d^4k}{(2\pi)^4} G(\tilde p_1 ,m_q) G(\tilde p_2 ,m_q) G(k,m_{e}) ,\label{eq:pert}
\end{align}
where $\tilde p_1 = p_1 + k$ and $\tilde p_2 = p_2 - k$, $m_q$ is the quark mass, and $m_{e}$ is the exchanged particle mass. Its dispersive representation reads
\begin{align}
F_0(s) = \frac{1}{\pi } \int_{s_t}^\infty \, ds' \, \frac{\Delta F_0(s')}{s' - s} \, , \label{eq:disp_rel}
\end{align}
where the discontinuity is computed from the diagram in~\cref{fig:modelpert} by putting the $q\bar{q}$ pair on shell according to the Cutkosky's cutting rules~\cite{Collins:1977jy,Eden:1966dnq}. This corresponds to the S wave of the $s$-channel partial wave projection of the \mbox{$q\bar q \to \pi \pi$} scattering amplitude  
\begin{align}
\Delta F_0(s') = & -\, \int \frac{d^4k}{(2\pi)^2} \frac{ \delta\left( \tilde p_1^{\, 2} - m_q^2\right) \, \delta\left( \tilde p_2^{\, 2} - m_q^2\right) }{ k^2 - m_{e}^2} \nonumber \\
= &\phantom{-} \int_{-1}^{1} dz \, t_{q\bar q \to \pi\pi}(s',t(s',z)) \, ,
\end{align}
where
\begin{align}
t_{q\bar q \to \pi\pi}(s',t(s',z))  = \frac{1}{8\pi} \, \frac{\rho_{q\bar{q}}}{s'\left( 1 - z \, \rho_{q\bar{q}} \, \rho_{2\pi}\right) + 2m_1^2 } \, , \label{eq:imf_6}
\end{align}
with \mbox{$m_1^2 = m_{e}^2 - m_\pi^2 - m_q^2$} and $z \equiv \cos \theta_s$, $\theta_s$  the $s$-channel scattering angle between one of the quarks and the pion it hadronizes into. The phase space reads \mbox{$\rho_i =\sqrt{ 1 - 4m_i^2 / s}$}.  After integrating over the scattering angle one finds for the discontinuity
$\Delta F_0(s')$,
\begin{align}
\Delta F_0(s') = \frac{1}{8\pi s' \rho_{2\pi} } \ln  \left|\frac{ s' \left(1 + \rho_{q\bar{q}}\rho_{2\pi}\right) + 2\, m_1^2 }{ s' \left(1 - \rho_{q\bar{q}}\rho_{2\pi}\right) + 2\, m_1^2} \right|  . \label{eq:imf_1}
\end{align}
We can now examine what~\cref{eq:imf_6} implies for the large $s$ behavior of~\cref{eq:disp_rel}. For $s \gg -t$, the quark exchange behaves as \mbox{$\simeq \left[m_{e}^2 -t(s',z)\right]^{-1}$} and \mbox{$t(s',z) \simeq s'\,(1-z)$}, so the integration over $z$ produces a $\ln (s')/s'$ contribution  and the $s'$ integration provides a second  $\ln s$, leading to
\begin{align}
F_0(s) \sim \frac{\ln^2s}{s}. \label{eq:pert_ref}
\end{align}

\subsection{Pion wave function effects} \label{sec:power_law}
The presented pointlike model lacks nonperturbative effects from the pion wave function. 
In the constituent quark model the pion can be described as a $q\bar{q}$ state. Therefore we can introduce a pion wave function by replacing the point-like quark propagators with constituent quark propagators. This is achieved by employing a dressed quark propagator which for simplicity, takes the form
\begin{align}
G(p,m) \to \left[G(p,m)\right]^n , \label{eq:power_sub}
\end{align}
where the $n>1$ subscript indicates the {\it softening power}. We have several softening options depending on how many and which propagators are softened. First we soften the two diagonal quark propagators (see~\cref{fig:modelpert}). The form factor $F_n(s)$ reads
\begin{align}
F_n(s) = \frac{1}{i} \int \frac{d^4k}{(2\pi)^4} \left[ G(\tilde p_1 ,m_q) \, G(\tilde p_2 ,m_q)\right]^n \, G(k,m_{e})   .\label{eq:power_law} 
\end{align}
Obviously, if $n=1$, the hard production $F_0(s)$ result is recovered. This simple model satisfies the condition of~\cref{eq:zero_moments} providing a form factor whose leading high energy behavior depends on $n$ according to (see~\cref{sec:powerlawall})
\begin{align}
F_n(s) \sim \frac{\ln s}{s^n}  \quad \text{for} \quad n > 1 . \label{eq:power_law_ref}
\end{align}

Therefore, this substitution with an arbitrary integer power $n$ cancels the first $n-1$ moments of the discontinuity of the form factor, yielding a softer pion form factor.

We can modify only the exchanged quark propagator in~\cref{fig:modelpert} following~\cref{eq:power_sub}
\begin{align}
F_{e,n}(s) = \frac{1}{i} \int \frac{d^4k}{(2\pi)^4} G(\tilde p_1 ,m_q) G(\tilde p_2 ,m_q) \left[ G(k,m_{e}) \right]^n  , \label{eq:vertical_power}
\end{align}
which at high energies behaves as (see~\cref{sec:powerlawex})
\begin{align}
F_{e,n}(s) \sim F_e(s) \sim \frac{\ln s}{ s} \quad \text{for} \quad n>1   . \label{eq:Fqn}
\end{align}

Hence, the asymptotic falloff of the form factor is not sensitive to the softening of the exchanged quark. This is expected, because the momentum  $k$ flowing through the vertical propagator in~\cref{fig:modelpert} can always be kept finite by having the quarks hadronize to pions with little transverse momentum, so that the hard momentum never flows through the exchanged quark line.

Finally, if we soften only one of the quark propagators connected to the photon vertex and the exchanged quark, we obtain
\begin{align}
F_{q,n}(s) = \frac{1}{i} \int \frac{d^4k}{(2\pi)^4} \left[ G(\tilde p_1 ,m_q) \right]^n  G(\tilde p_2 ,m_q) \,  G(k,m_{e}) \,  . \label{eq:f1}
\end{align}

We find that the high energy behavior does not depend on $n$ (see~\cref{sec:powerlawone})
\begin{align}
F_{q,n}(s)  \sim F_{q}(s)  \sim \frac{1}{s} \quad \text{for} \quad n > 1 . \label{eq:power_law1}
\end{align}

This result is a consequence of the large momentum flowing through the perturbative propagator $G(\tilde p_2 ,m_q)$ which leads to a nonvanishing 0$^\text{th}$ moment in~\cref{eq:zero_moments}. Therefore, the first term in the~\cref{eq:disp_exp} expansion contributes, leading to a $s^{-1}$ asymptotic behavior.

These results show why the notion of a pion wave function should not be associated exclusively with the \mbox{quark-pion} vertex function, which is a function of $t$ only, but with the intermediate state in the pion channels. These intermediate states contain the exchanged quark and one of the quark lines connected to the photon. Therefore a physical model with soft pion wave functions, in a dispersive representation,  can be realized through modification of the quark propagators. Such propagators emerge for example from calculations based on the QCD Dyson-Schwinger equations~\cite{Roberts:2012sv}.

Given that pions are soft, to bring  the power behavior of the form factor back to \mbox{$\sim s^{-1}$} as shown above, it is sufficient to screen only one of the two wave functions.  This  is the role of the one gluon exchange, which effectively short circuits one of the pion wave functions. 

\subsection{One gluon exchange}\label{sec:gluon}
\begin{figure}
\includegraphics[width=0.4\textwidth]{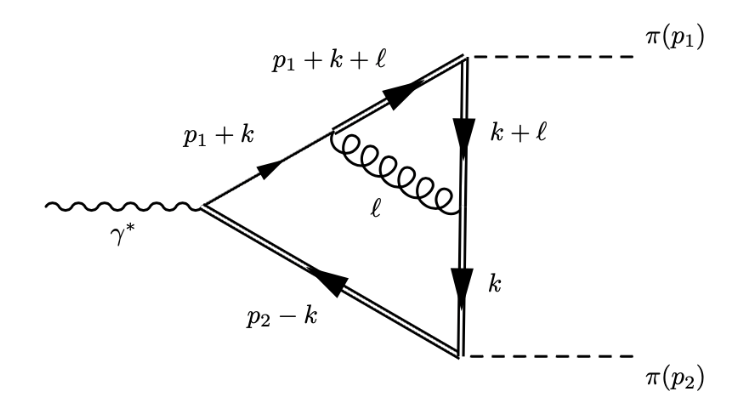}
\caption{Gluon exchange at a quark-pion interaction vertex. Propagators softened according to~\cref{eq:power_sub} are shown as double lines.} \label{fig:gluon_loop}
\end{figure}

As the energy of the photon increases, the spatial resolution of the photon becomes fine enough that the individual gluon exchange at the pion vertex becomes relevant. Therefore, the gluon exchange in~\cref{fig:gluon_loop} contributes to the high energy behavior of the form factor.

In the dispersive representation, the  quark-antiquark intermediate state  produced by the hard photon evolves to the $2\pi$ state. This evolution is  soft resulting in a form factor that  falls off rapidly with $s$. Emission of a single gluon from either quark followed by the gluon dissociation into a $q\bar q$ pair can be hard and therefore will dominate the form factor at large $s$. The time between gluon emission and decay to $q\bar q$ pair scales like $1/s$ giving  the leading power behavior of the form factor. The residual interactions between two $q\bar q$ are soft and can be represented by the soften propagators. 
    
If we combine this gluon exchange and the soft quark propagators that account for the pion wave function in~\cref{eq:power_law}, we obtain
\begin{align}
F_{g,n}(s) = \frac{1}{i} \int \frac{d^4k}{(2\pi)^4} G(\tilde p_1,m_q) \left[G(\tilde p_2,m_q) G(k,m_{e})\right]^n \, I_g, \label{eq:gluon_model}
\end{align}
where
\begin{align}
I_g = \frac{1}{i} \int \frac{d^4\ell}{(2\pi)^4} G(\ell,0) \left[G(k+\ell,m_e) G(\tilde p_1 + \ell,m_q) \right]^n ,
\end{align}

is the integral over the softened quark-gluon-quark triangle. Numerical and analytical (see~\cref{sec:gluonloop}) results for~\cref{eq:gluon_model} show that \mbox{$s F_{g,n}(s) \to \text{constant}$} regardless of $n$ for the $s\to\infty$ limit. The results show that the introduction of the gluon loop effectively screens the form factor from the softening effects of the nonperturbative power law modification leading to a 
\begin{align}
F_{g,n}(s) \sim F_g(s) \sim \frac{1}{s}  ,\label{eq:gluon}
\end{align}
behavior. Which, when the running of the strong coupling is incorporated, provides the expected pQCD behavior.

\section{Reggeized model}\label{sec:ReggeizedModel}
At high energies, the contribution of the $t$-channel exchange to the $q\bar q \to \pi\pi$ process is expected to be described by the corresponding Regge trajectory~\cite{Fadin:1977jr,Strikman:2007nz} instead of a single quark exchange. This effectively describes multiparticle production at high energies and introduces energy dependence to $t_{q\bar q \to \pi\pi}$ which otherwise would be only a function of the momentum transferred between the quark and the pion as in~\cref{eq:pert}. Regge theory provides a rigorous description for the amplitude at forward angles, which, at high energies, accounts for almost all the contribution to the cross section. Hence, once the angular dependence is integrated, the asymptotic behavior of the $q\bar{q}\to \pi \pi $ amplitude in~\cref{fig:modelpert} can be described by the leading Regge trajectory $\alpha(t)$ of the exchanged particle as follows 
\begin{align}
 t_{q \bar{q} \to \pi \pi}  \sim \beta(t) s^{\alpha(t)}. \label{eq:regge_amp}
\end{align}
As discussed in~\cref{sec:pert}, asymptotically, the form factor is dominated by the $s \gg t$ region. Therefore, the finite $t$ dependence of the Regge residues $\beta(t)$ does not affect the asymptotic behavior and can be safely ignored. Once the Regge trajectory $\alpha(t)$ is determined, the pion form factor can be calculated by replacing the exchanged quark propagator in~\cref{eq:pert} by
\begin{align}
G(k,m_{e}) \to s^{\alpha(k^2)}  , \label{eq:regge_prop}
\end{align}
leading to
\begin{align}
F_{R}(s) = \frac{1}{i} \int \frac{d^4k}{(2\pi)^4} G(\tilde p_1,m_q) \, G(\tilde p_2,m_q) \, s^{\alpha(k^2)} . \label{eq:reggeformfactor}
\end{align}

The form factor is straightforwardly calculated using the dispersive integral of~\cref{eq:dicFormFactorGeneral} with the discontinuity given by
\begin{align}
\Delta F_R(s) = - \, \int \frac{d^3k}{(2\pi)^2} \frac{ \delta(|\vec{k}| - k^*) }{|\vec{k}| \sqrt{s}}  s^{\alpha\left(-(\vec{k} - \vec{p}_1)^2\right)}  ,
\end{align}
where $k^* =\sqrt{s/4-m_q^2}$, which in the $s \gg t$ regime becomes 
\begin{align}
\Delta F_R(s) \simeq - \int_{-1}^{1} dz \,s^{\alpha(t)} . \label{eq:im_regge}
\end{align}
The only unknown piece is the functional form of the Regge trajectory, which is an analytical function with poles at the angular momenta of the particles that belong to the trajectory and has no left-hand cuts~\cite{Collins:1977jy}. Additionally, it has to satisfy 
\begin{align}
\lim_{t\to-\infty} \Re \alpha(t) = -1  , \label{eq:reggelimit}
\end{align}
as the physics condition required by perturbation theory for scalar particles~\cite{Blankenbecler:1973kt}. \cref{sec:scalar_qrt} provides the self-consistent calculation of the Regge trajectory.

The asymptotic behavior can be established analytically and this is done in \cref{sec:reggeizedffappendix} where all the details are provided. The asymptotic behavior depends exclusively on the value of the intercept of the Regge trajectory, \ie \mbox{$\alpha(t=0)$}. We find three scenarios: (a) \mbox{$\alpha(0)\le 0$}; (b) \mbox{$0< \alpha(0) < 1$}; and (c) \mbox{$\alpha(0) \ge 1$}. If \mbox{$\alpha(0) \ge 1$} the integral in~\cref{eq:dicFormFactorGeneral} does not converge and it needs to be subtracted. The obtained Regge trajectory corresponds to case (a), so the asymptotic behavior is
\begin{align}
F_R(s) \sim \frac{\ln s}{s} \, . \label{eq:reggeff}
\end{align}
If no additional effects are considered, Eq.~\cref{eq:reggeff} is independent of any trajectory parameters, therefore, the asymptotic behavior for any trajectory with $\alpha(0) \le 0$ is universal. In case (b), the high energy behavior is \mbox{$\tilde{F}_R(s) \sim s^{\alpha(0)} \slash s \ln s$}, as obtained in \cref{eq:alfa1}. In case (c), we have to subtract the dispersion relation to make it converge. Both cases (b) and (c) are harder than the pointlike model and are, hence, unphysical. The phenomenological Regge trajectory employed in \cite{Gorchtein:2011vf} belongs to case (b).

As shown in \cref{sec:gluon}, the exchange of one gluon hardens the soft quark propagator.  Hence, the combination of hard and soft quark propagators represents the effects of a gluon exchange that screens one of the two pion wave functions, and in the presence of Reggeization, the form factor reads
\begin{align}
F_{R,n}(s) = \frac{1}{i} \int \frac{d^4k}{(2\pi)^4} G(\tilde p_1,m_q)  \left[ G(\tilde p_2,m_q) \right]^n \, s^{\alpha(k^2)} . \label{eq:softened_regge}
\end{align}

Once the Regge trajectory is determined, the form factor can be numerically computed. We do this calculation for the  $\alpha(0)\le0$ case. More details are given in~\cref{sec:regsoft}. We find that the asymptotic behavior of $F_{R,n}$ is independent of the softening power $n$, as shown in~\cref{sec:power_law}, and a consequence of the large momentum flowing through the perturbative propagator $G(\tilde p_1 ,m_q)$ which leads to a nonzero 0$^\text{th}$ moment in~\cref{eq:zero_moments}. Therefore, the first term in the~\cref{eq:disp_exp} expansion contributes, leading to
\begin{align}
F_{R,n} \sim \frac{1}{s} ,
\end{align}
which is compatible with the expected pQCD behavior once the running of the strong coupling is incorporated.

\section{Summary and conclusions}\label{sec:conclusions}
\begin{table}
\caption{Summary of the high energy behavior for the different models of the electromagnetic pion form factor with \mbox{$n>1$} the softening power. The mechanisms in the top panel are compatible with the pQCD asymptotics, once the running of $\alpha_s$ is added. The mechanisms listed in the bottom panel are incompatible with the pQCD expectation.} \label{tab:summary}
\begin{ruledtabular} \begin{tabular}{lr}
Softened single quark        &  $F_{q,n} \sim 1 \slash s $ \\
Gluon exchange and softening  &  $F_{g,n} \sim 1 \slash s $ \\
Reggeized and softening,  $\alpha(0)\le 0$    &  $F_{R,n} \sim 1\slash s $ \\
\hline
Pointlike                   &  $F_0 \sim \ln^2 s \slash s $ \\
Softened quarks             &  $F_n \sim \ln s \slash s^n $ \\
Softened exchange            &  $F_{e,n} \sim \ln s \slash s $ \\
Reggeized, $\alpha(0)\le 0$  &  $F_{R}   \sim \ln s \slash s $ \\
Reggeized, $0<\alpha(0)< 1$     &  $\tilde{F}_{R} \sim s^{\alpha(0)} \slash s \ln s $
\end{tabular} \end{ruledtabular} 
\end{table}

The discrepancy between the asymptotic pQCD prediction and the experimental data for the electromagnetic pion form factor shows that experiments have not reached the perturbative regime yet. This calls for nonperturbative models that can transition between soft and hard processes and are applicable to the current experimental data. These models should be compatible with the pQCD prediction at higher energies. To provide a path to connect the perturbative and nonperturbative regimes, we have explored what conditions a Regge exchange model has to fulfill, in order to be consistent with the pQCD result. 

We have used the framework of dispersion relations to ensure that the fundamental $S$-matrix constraints are fulfilled. We have studied the high-energy asymptotic behavior of the electromagnetic pion form factor in  simplified models to access the role of different microscopic effects. The form factor has been modeled as a triangle diagram where the photon splits into a $q\bar{q}$ pair that exchanges a (Reggeized) quark to produce the two pions. 

\cref{tab:summary} summarizes our findings for the asymptotics of the various mechanisms explored. Nonperturbative effects such as the pion wave function have been incorporated by softening the $q\bar{q}$ pair production. This modification changes the asymptotic leading power, resulting in a faster fall-off of the form factor.  However, the introduction of gluon loops effectively screens the form factor from the softening effects of the pion wave function. The constructed model simultaneously accounts for the pion wave function, gluonic exchanges, and quark Reggeization.

We also find that the inclusion of nonperturbative effects in the quark exchange and its Reggeization only affects the leading logarithmic correction. For the Reggeized case, we find three different asymptotics depending on the intercept of the trajectory, setting a strict constraint: \mbox{$\alpha(0)\le 0$}. Having the correct asymptotic behavior requires the combination of softening and Reggeization. The transition  at $\alpha(0)=0$ corresponds to the appearance of a spin-$0$ massless particle in the spectrum, which is not realized in nature. Nevertheless, it may be worth exploring the possibility of an enhancement of the trajectory within the context of Reggeized spin-$1/2$ quarks.

\acknowledgments
C.F.R. thanks Sean Dobbs for his insight on the \mbox{CLEO-c} data. This work was supported by the U.S. Department of Energy contract \mbox{DE-AC05-06OR23177}, under which Jefferson Science Associates, LLC operates Jefferson Lab, and \mbox{DE-FG02-87ER40365}, by the U.S. National Science Foundation Grant \mbox{No.~PHY-2310149}, and by the Spanish Ministerio de Ciencia, Innovación y Universidades \mbox{(MICIU)} Grant \mbox{Nos.~PID2020-118758GB-I00} and \mbox{CNS2022-136085}. CFR is supported by Spanish Ministerio de Ciencia, Innovación y Universidades \mbox{(MICIU)} under Grant \mbox{No.~BG20/00133}. V.M. acknowledges support as a Serra Húnter Fellow. D.W. was supported in part  by DFG and NSFC through funds provided to the Sino-German CRC 110 ``Symmetries and the Emergence of Structure in QCD'' (NSFC Grant \mbox{No.~12070131001}, DFG Project-ID \mbox{196253076}). This work contributes to the aims of the U.S. Department of Energy ExoHad Topical Collaboration, contract \mbox{DE-SC0023598}.

\appendix
\section{Loop integrals for the softened triangle diagram} \label{sec:powerlaw}
\subsection{Quark propagators softened} \label{sec:powerlawall}
To proceed with~\cref{eq:power_law} we use Feynman parameters for positive integer $n$. Following the conventions of~\cite{Srednicki:2007qs}, the result is 
\begin{align}
F_n(s) = A_n \int \, dx \, dy \, \frac{ \left[x\,y \right]^{n-1} \,(1-x-y)}{ \mathcal{D}^{2n-1} } \, , \label{power_law}
\end{align}
where 
\begin{align}
\mathcal{D} = & \, xys - (x+y)^2 m_\pi^2  \nonumber \\
& + (x+y)(m_\pi^2 + m_{e}^2 - m_q^2) - m_{e}^2 \, , \\
A_n = &  \frac{\Gamma(2n-1)}{ (4 \pi)^2 \left[ \Gamma(n)\right]^2} \, .
\end{align}

For large $s$ the integral will be dominated in the region $x,y \sim s^{-1}$. Perform the integral in $x$ and expand around large $s$ with $x = s^{-1}$, and then do the same with $y$. From this, the leading $s$ behavior is obtained to be~\cref{eq:power_law_ref}. 

\subsection{Only the exchanged particle propagator softened}\label{sec:powerlawex}
The result for softening only the exchanged quark propagator is found replacing one factor of $n$ in~\cref{eq:power_law} by a factor of $1$, replacing $A_n$ with $1/16\pi^2$, and removing one Feynman variable $x$ (or $y$) from the numerator
\begin{align}
F_{e,n}(s) = \frac{1}{16\pi^2} \int \, dx \, dy \, \frac{x^{n-1}}{ \mathcal{D}^{n} } \, . \label{eq:single_soft_quark}
\end{align}

At large $s$ the dominant contribution to this integral is found in the region $y\sim s^{-1}$ and $x\sim 1$. If one integrates~\cref{eq:single_soft_quark} and expands the result in large $s$ in this region of $x$ and $y$, the form factor is found to behave as~\cref{eq:Fqn}.

To see the effect of altering the power of the vertical propagator as in~\cref{eq:vertical_power} we use the dispersive technique of~\cref{sec:pert}. According to the delta functions attained by putting the quarks on shell, the discontinuity of the form factor becomes
\begin{align}
\Delta F_n(s) = \frac{  \rho_{q\bar{q}}}{8 \pi } \int \frac{dz}{\left[s \left( z \rho_{2\pi}\,  \rho_{q\bar{q}} - 1\right)\slash 2  + m_\pi^2 + m_q^2 - m_{e}^2 \right]^n} \, .
\end{align}
where $z$ is the cosine of the polar angle of the loop momentum in spherical coordinates. This integral peaks when \mbox{$z\, \rho_{2\pi}\, \rho_{q\bar{q}} - 1 \sim s^{-1}$} so that the denominator is not large. In this region, the $s$ dependence of the denominator is completely canceled, and it is clear that the leading behavior of form factor will not pick up any additional $s$ dependence from the integration in~\cref{eq:dicFormFactorGeneral}.

\subsection{Only one quark propagator softened} \label{sec:powerlawone}
To study the behavior of the form factor with one softened quark propagator we turn again to Feynman parameters
\begin{align}
F_{q,n}(s) = B_n \, \int \, dx \, dy \, \frac{ x^{n-1}\,(1-x-y)}{ \mathcal{D}^{n}} \label{eq:fqn} \, , 
\end{align}
where \mbox{$B_n = 1 / \left[(4\pi)^2\Gamma(n)\right]$}. The dominant contribution to this integral comes from the region \mbox{$y\sim s^{-1}$} and \mbox{$x \sim 1$}. In considering the dependence of $\mathcal{D}$ and of the numerator of~\cref{eq:fqn} on the Feynman parameters, it is clear that in this region the form factor with one softened quark will fall off at most as fast as $s^{-1}$.

\section{Insertion of a gluon exchange} \label{sec:gluonloop}
We now examine the large $s$ behavior of the gluon loop. Using Feynman parameters the gluon loop can be made to look like a propagator with dependence on the loop momentum $k$
\begin{align}
I_g =  \tilde B_n \int dx_1 \, dx_2 \, \frac{ (x_1 x_2)^{n-1} \; G(\hat k,\hat m )^{2n-1} }{ \left[ (x_1+x_2) x_3  \right]^{2n-1}}  \,  ,
\end{align}
where $x_3 = 1-x_1-x_2$ and
\begin{subequations}
\begin{align}
\tilde B_n &= \frac{\Gamma(2n-1)}{\Gamma(n)} \, B_n ,\\
\hat k &= k + \frac{x_1}{x_3} \, p_1  , \\
\hat m^2 &= \frac{ x_1(1 - x_2 - 2x_1x_3)m_\pi^2 - x_1 x_3 m_q^2 + x_2 x_3 m_e^2 }{x_3^2 (x_1 + x_2)} \, .
\end{align}
\end{subequations}

Combining this with the other propagators in~\cref{eq:gluon_model} and using Feynman parameters $y_i$ we find
\begin{widetext}
\begin{align}
F_{g,n}(s) &=  \frac{\tilde B_n}{i} \frac{\Gamma\left(3n+1\right)}{\Gamma\left(2n-1\right)\Gamma\left(n\right)} \int \frac{d^4k \, dx_1 \, dx_2 \, \left(x_1 x_2\right)^{n-1}}{\left[\left(x_1+x_2\right)\left(1-x_1-x_2\right)\right]^{2n-1}}  \nonumber \\
 & \times \frac{dy_1 \, dy_2 \, dy_3 \, y_2^{n-1} \, y_3^{2n-2} }{\left[ y_1\left(\hat p_1^2 - m_q^2\right) + y_2\left(\hat p_2^2 - m_q^2\right) + y_3\left(\hat k^2 - \hat m^2\right) + y_4\left( k^2 - m_{e}^2\right) \right]^{4n}}  ,
\end{align}
\end{widetext}
where it is understood that \mbox{$y_4 = 1 - y_1 - y_2 - y_3$}. We complete the square in the loop momentum, Wick rotate, and perform the loop momentum integration to get
\begin{align}
F_{g,n}(s) = C_n  \int  \frac{dX}{\mathcal{G}^{4n-2}} \frac{ \, (x_1 x_2 y_2 y_4)^{n-1} \, y_3^{2n-1}}{\left[(x_1+x_2)x_3\right]^{2n-1} }  ,
\end{align}
where $y_4 = 1-y_1-y_2-y_3$ and
\begin{align}
\mathcal{G} =& y_2 (y_1 - \frac{x_1 y_3}{x_3}) s + b  , \\
C_n =& \frac{\Gamma(4n-2)}{\left( 16\pi^2 \Gamma(n)^2\right)^2}  ,
\end{align}
with $dX \equiv dx_1 dx_2 dy_1 dy_2 dy_3$ and $b$ that depends on the Feynman parameters and is independent of $s$. According to the form of $\mathcal{G}$, the form factor will be very small unless the coefficient of $s$ is of the order $s^{-1}$ or smaller in the large $s$ limit. The leading order behavior will come from the integration region \mbox{$x_1 \sim x_2 \sim y_1 \sim s^{-1}$} and $y_2 \sim y_3 \sim 1$. From here it can be shown that there will remain a term with leading behavior \mbox{$F_{g,n}(s) \sim s^{-1}$}.

\section{Scalar Regge trajectory}\label{sec:scalar_qrt}
\begin{figure}
\includegraphics[width=0.45\textwidth]{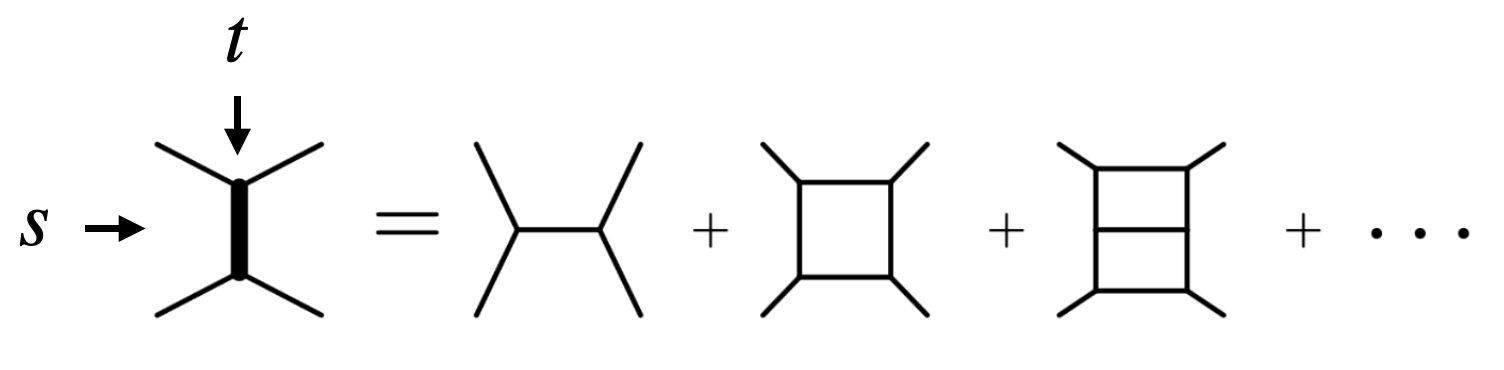}
\caption{Construction of the Reggeon exchange amplitude $A(s,t)$ as the sum of all ladder diagrams with particles exchanged in the $t$ channel.} \label{fig:ladder_sum}
\end{figure}

The starting point for our model of the Regge trajectory is the $t$ channel Reggeon exchange amplitude $A(s,t)$  where, for simplicity, all the incoming and outgoing particles are assumed to be equal-mass scalars, \ie \mbox{$m\equiv m_q=m_\pi$} and the exchanged Reggeon has mass $m_{e}$. $A(s,t)$ is the sum of all ladder diagrams with particles exchanged in the $t$ channel as depicted in~\cref{fig:ladder_sum}. It was shown in~\cite{Gell-Mann:1964aya} that such resummation produces an amplitude that behaves as expected in~\cref{eq:regge_amp}.

The first term in the sum of ladder diagrams is a tree-level amplitude from a simple quark exchange, which in the center-of-mass frame can be expanded in $t$-channel partial waves,
\begin{align}
A^1(s,t) = \frac{g^2}{m_{e}^2-s(t,z_t)}=\sum_{\ell=0}^\infty (2\ell+1) A^1_\ell(t) P_\ell(z_t), \label{eq:tree}
\end{align}
where $g$ is the pion-quark-Reggeon coupling, $P_\ell$ are the Legendre polynomials of the first kind, and  \mbox{$z_t=\cos \theta_t=1 +  2 s \slash \left( t - 4m^2 \right)$} with $\theta_t$ the $t$-channel scattering angle. We can use  unitarity to reconstruct the ladder diagram in terms of the tree level  $t$ partial wave. The partial wave is
\begin{align}
A_\ell^1(t) =& \frac{1}{2} \int_{-1}^1 d z_t \, A^1(s,t) P_\ell(z_t) \nonumber \\
= & \frac{g^2}{2} \int_{-1}^1 d z_t\, \frac{ P_\ell(z_t)}{ m_{e}^2 - s(t,z_t) } \nonumber \\ 
= &  \frac{g^2}{2k^2}  (-1)^\ell Q_\ell(z_0) \, , \label{eq:tree_level}
\end{align}
where \mbox{$z_0 = 1 + 2 m_{e}^2 \slash \left( t - 4m^2 \right)$} is the value of $z_t$ for which \mbox{$s=m_{e}^2$},  \mbox{$k^2 = \left( t-4m^2\right)\slash 4$}, and $Q_\ell$ are the Legendre functions of the second kind. The $Q_\ell$'s can be expanded using hypergeometric functions
\begin{widetext}
\begin{align}
Q_\ell(z_0) =& \frac{\sqrt{\pi}}{(2z_0)^{\ell+1}} \frac{\Gamma(\ell+1)}{\Gamma(\ell+3/2)}
\, _2F_1\left( \ell/2+1,\ell/2+1/2;\ell+3/2;1/z_0^2\right) \nonumber \\
= & \, \frac{\sqrt{\pi}}{2^{\ell+1}} \left(\frac{\nu}{\nu+\bar{r}}\right)^{\ell+1}
\sum_{n=0}^\infty \frac{\Gamma(2n+ \ell+1)}{\Gamma(n+\ell+3/2) \: n!} \left(\frac{\nu}{2(\nu+\bar{r})}\right)^{2n}  =  \, \nu^{\ell+1} H(\ell,\nu) \, ,\label{eq:Qexpanded}
\end{align}
with
\begin{align}
H(\ell,\nu) = \frac{1}{2^{\ell+1}} \frac{\sqrt{\pi}}{\left(\nu+\bar{r} \right)^{\ell+1}} \sum_{n=0}^\infty \frac{\Gamma(2n+ \ell+1)}{\Gamma(n+\ell+3/2) \: n!} \left(\frac{\nu}{2(\nu+\bar{r})}\right)^{2n} \, , \label{eq:G}
\end{align}
\end{widetext}
where $\nu = t/4m^2-1$ and  $\bar{r} = m_{e}^2 \slash 2m^2$. The series converges because $\nu,\, \bar{r}>0$, implying \mbox{$\nu/(\nu+\bar{r})<1$.}  We note that $H(\ell,\nu)$ has a leading pole  in the complex angular momentum plane at $\ell=-1$. We can rewrite~\cref{eq:tree_level} as 
\begin{align}
A_\ell^{1}(\nu) =& \frac{g^2}{2m^2} \frac{\nu^\ell}{\ell+1} \, (\ell+1)\,(-1)^\ell  \, H(\ell,\nu) \nonumber \\
\equiv& \frac{g^2}{2m^2} \frac{\nu^\ell}{\ell+1} \, \hat{N}(\ell,\nu) \, , \label{eq:A1eqN}
\end{align}
where the leading pole has been made explicit and $\hat N(\ell,\nu)$ is finite and nonzero at the pole. Since $\ell<0$, this pole does not correspond to a physical particle in the $t$ channel.  In the following, we assume this property will not change after unitarization. Physically this corresponds to a weak coupling, \ie the one-boson exchange is not strong enough to bind two scalars. In the case of spinors, however, the leading pole will occur at $\ell = 1/2$, in which case the Regge amplitude would corresponds to the exchange of a quark in the $t$-channel. From the partial wave amplitude in~\cref{eq:tree_level} we can construct higher order amplitudes such as 
\begin{align}
 A^2_\ell(\nu) = \frac{1}{\pi} \int_0^\infty d\nu' \frac{A^1_\ell(\nu') \sigma (\nu') A^1_\ell(\nu')}{ \nu' - \nu },
\end{align}
and so on, where 
\begin{align}
\sigma(\nu) = \frac{\lambda^{1/2}(t,m^2,m^2)}{ 16 \pi t}= \frac{1}{16\pi}\sqrt{\frac{\nu}{\nu+1}}\, .
\end{align}

The full amplitude partial waves can be approximated using the $N/D$ formalism~\cite{Chew:1960iv} defining
\begin{align}
A_\ell(\nu) =\frac{ N_\ell(\nu)}{ D_\ell(\nu)}= \frac{A^{1}_\ell(\nu)}{1-\frac{\nu}{\pi}\int_0^\infty d\nu' \sigma(\nu')\frac{A^{1}_\ell(\nu')}{\nu'(\nu'-\nu)}} \, ,
\end{align}
which can  be rewritten as a Regge amplitude in the form of a simple Regge pole using \cref{eq:A1eqN}
\begin{align}
A_\ell(\nu) = \frac{g^2}{2m^2} \frac{\nu^\ell \hat{N}(\ell,\nu)}{\ell-\hat{D}(\ell,\nu)}
= \frac{\beta(\nu)}{\ell -\alpha(\nu)} \, , \label{eq:Apole}
\end{align}
where $\hat{N}$ is defined in \cref{eq:A1eqN} and
\begin{align}
\hat{D}(\ell,\nu) = \ell+1 -\frac{g^2}{2m^2}  \frac{\nu}{\pi} 
\int_0^\infty d\nu' \nu'^\ell \, \frac{\sigma(\nu')\hat{N}(\ell, \nu')}{\nu'(\nu'-\nu)} \, .
\end{align}
Hence,
\begin{align}
\alpha(\nu) = \hat{D}(\alpha(\nu),\nu) \quad  \textrm{and} \quad \Im \alpha(\nu) = \sigma(\nu)\beta(\nu) .
\end{align}

\begin{figure}
\includegraphics[width=0.48\textwidth]{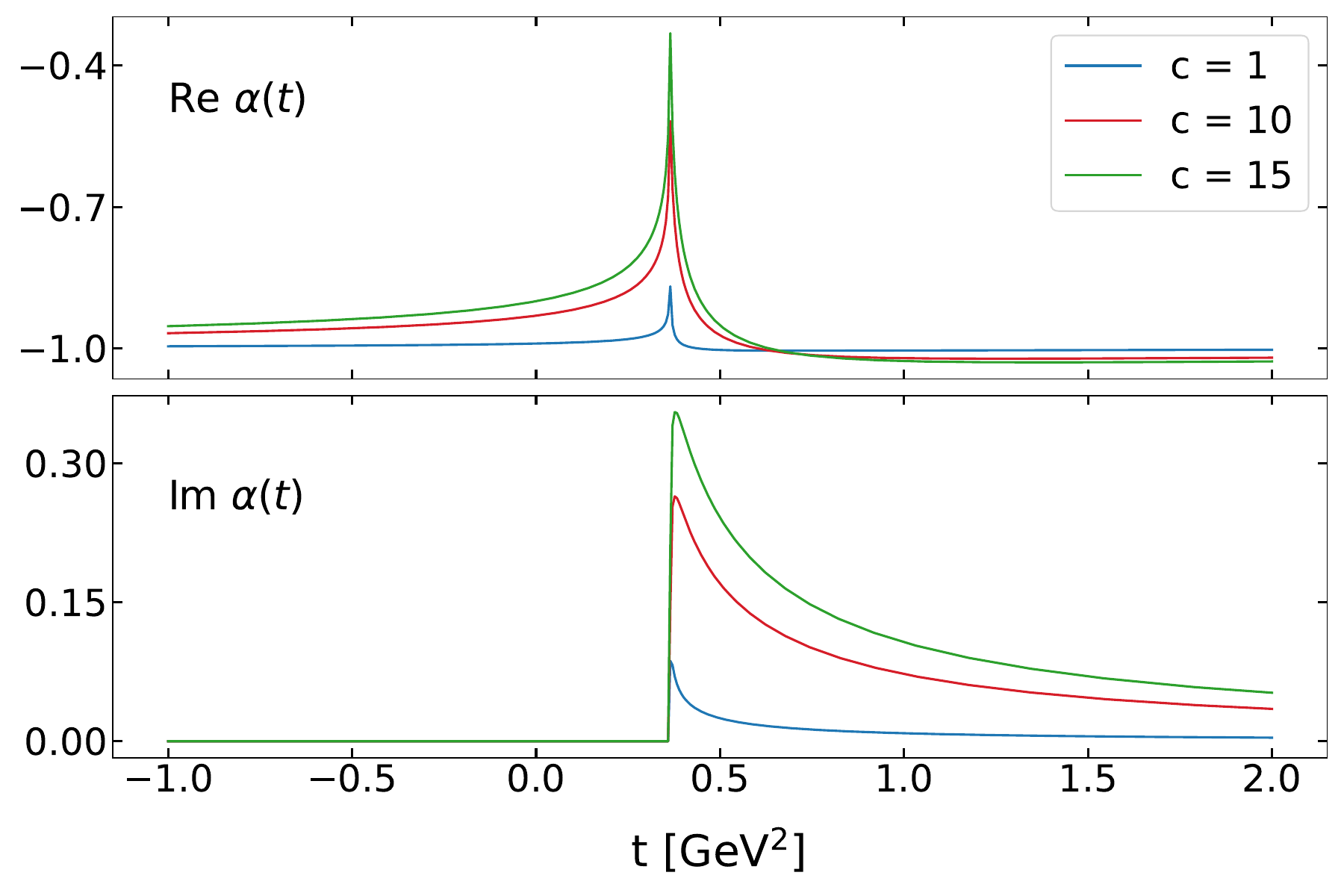}
\caption{Example of the calculation of the Regge trajectory $\alpha(t)$ using~\cref{eq:G,eq:beta,eq:im_a,eq:re_a,eq:restriction} for \mbox{$c=\{1,10,15\}$} and $\bar{r}=1$.}
\label{fig:regge_traj}
\end{figure}

Following~\cref{eq:Apole}, $\beta(\nu)$ can be written
\begin{align}
\beta(\nu) = \gamma(\nu) \nu^{\alpha(\nu)} |\hat{N}(\alpha(\nu),\nu)| \text{e}^{i \phi(\nu)} .
\end{align}
where \mbox{$\ell \to \alpha(\nu)$} and \mbox{$\phi(\nu) = \arg H(\alpha(\nu),\nu)$}. The residue satisfies \mbox{$\beta(\nu^*) = \beta^*(\nu)$} in the complex $\nu$ plane cut along the real axis for $\nu>0$ (Schwarz reflection principle). Hence, $\beta(\nu)$ is a real function above threshold and $\arg \beta(\nu)=0$, which implies
\begin{align}
\arg \gamma(\nu)=-\phi(\nu)-\Im \alpha(\nu) \ln \nu  \, .
\end{align}
and
\begin{widetext}
\begin{align}
\beta(\nu) = c \, |H(\alpha(\nu),\nu)| \, \exp\left[ \Re\alpha(\nu)\, \ln\nu- \frac{\nu+1}{\pi} \textrm{PV}\int_0^\infty d\nu' \frac{\phi(\nu^\prime)+\Im\alpha(\nu')\ln\nu'}{(\nu'+1)(\nu'-\nu)} \right]  ,  \label{eq:beta}
\end{align}
\end{widetext}
where $\textrm{PV}$ stands for Cauchy principal value and \mbox{$c = \gamma_0 g^2/2m^2$} with $\gamma_0$ an unknown constant to be determined~\cite{Londergan:2013dza,Pelaez:2017sit}. The unitarity of the amplitude in~\cref{eq:Apole} allows to compute the Regge trajectory $\alpha(\nu)$ from $\beta(\nu)$
\begin{align}
\Im \alpha(\nu) &= \sigma(\nu) \beta(\nu)  \label{eq:im_a} \, , \\
\Re \alpha(\nu) &= a + \frac{\nu+1}{\pi} \text{PV}\int_0^\infty d\nu' \frac{\Im \alpha(\nu')}{ (\nu' + 1) (\nu' - \nu) } , \label{eq:re_a} 
\end{align}
with $a$ chosen such that it satisfies the physics condition required by perturbation theory for scalar particles~\cite{Blankenbecler:1973kt} in~\cref{eq:reggelimit}. Consequently
\begin{align}
a = -1 + \frac{1}{\pi} \int_0^\infty d\nu' \frac{\Im \alpha(\nu')}{\nu' + 1} . \label{eq:restriction}
\end{align}
Therefore, given the values of $c$ and $\bar{r}$ it is possible to obtain iteratively the Regge trajectory $\alpha(\nu)$ using~\cref{eq:G,eq:beta,eq:im_a,eq:re_a} while imposing~\cref{eq:restriction}. 

We explored a wide range of values of these two parameters and found that the trajectory depends weakly on the value of $\bar{r}$, \ie the masses of the particles involved. Convergence of the iterative equations occurs well within ten iterations for $c \leq 15$. In~\cref{fig:regge_traj} we show the Regge trajectory for $\bar{r}=1$ and several values of $c$. We find that $c$ determines the size of the peak shown in~\cref{fig:regge_traj} and has a mild impact in the form factor.

\section{Asymptotic behavior of the Reggeized form factor} \label{sec:reggeizedffappendix}
We are interested in the $s\to \infty$ behavior of the integral in~\cref{eq:im_regge}
\begin{align}
\Delta F_R(s) = &- \int_{-1}^{1} dz\, s^{\alpha(t)} = - \int_{-1}^{1} dz\, s^{-1+h(t)}  \nonumber \\
 = &- \frac{1}{s}\int_{-1}^{1} dz\, \text{e}^{h(t)\ln s} \, .
\end{align}

We split the integral into two pieces to identify the two contributions to the large $s$ behavior
\begin{align}
\Delta F_R(s) = -\frac{1}{s} \int_{-1}^{1-\varepsilon} \, dz \, e^{h(t) \ln s} - \frac{1}{s} \int_{1-\varepsilon}^1 \, dz \, e^{h(t) \ln s} \, ,
\end{align}
with \mbox{$\varepsilon \sim 1/s$}. In the first integral all values of $z$ are finitely large, and the tendency of \mbox{$h(t) \to 0$} as \mbox{$t\to-\infty$} leads to the result
\begin{align}
\frac{1}{s} \int_{-1}^{1-\varepsilon} \, dz \, e^{h(t) \ln s} = \frac{2-\varepsilon}{s} \simeq \frac{2}{s}  .
\end{align}

In the second region of integration, small values of $\varepsilon$ will lead to small values of \mbox{$t(s,z) \sim -s(1-z)$}. Thus we can expand $h(t)$ around $t=0$, resulting in 
\begin{equation}
\frac{1}{s} \int_{1-\varepsilon}^1 \, dz \, e^{h(t) \ln s} = \frac{1}{s} \int_0^\varepsilon \, du \, e^{ \left[ h(0) - s\,u\,h'(0) \right] \ln s} \, ,
\end{equation}
where we have made the change of variable $u = 1-z$. We note that \mbox{$h'(0) = \alpha'(0)$}. This integral is straightforward to evaluate
\begin{align}
\frac{1}{s} \int_{1-\varepsilon}^1 \, dz \, e^{h(t) \ln s} = \frac{s^{-2 + h(0)}}{\alpha'(0) \ln s} \left[ 1 - s^{-\alpha'(0)} \right] \, . \label{eq:high_z_regge}
\end{align}

For $t<0$ we observe that $\alpha'(t) > 0$, and so for large $s$ we expect~\cref{eq:high_z_regge} to be dominated by the first term in brackets. The full result at large $s$ is then
\begin{align}
\Delta F_R(s) = -\frac{1}{s} \left( 2 + \frac{s^{\alpha(0)}}{ \alpha'(0) \, \ln s} \right) \, . \label{eq:c6}
\end{align}

If \mbox{$\alpha(0) \le 0$ ($> 0$)}  the leading behavior is the first (second) term. Considering that, as shown in~\cref{fig:regge_traj}, \mbox{$\alpha(0)<0$}, the first term dominates and using~\cref{eq:disp_exp},  we obtain that the leading behavior of the form factor is
\begin{align}
F_{R}(s) \sim \frac{\ln s}{s} \, .
\end{align}

For a Regge trajectory with $\alpha(0)> 0$, the discontinuity at high energies reads
\begin{align}
\Delta \tilde{F}_R(s) \simeq - \frac{s^{\alpha(0)}}{s \ln s} \, ,
\end{align}
and using~\cref{eq:disp_exp} we find for $0<\alpha(0)<1$
\begin{align}
\tilde{F}_R(s) \simeq \frac{\text{Ei} \left[ \alpha(0) \ln s \right]}{s} \sim \frac{s^{\alpha(0)}}{ s \ln s} \, . \label{eq:alfa1}
\end{align}
If $ \alpha(0) \ge 1$ the integral in~\cref{eq:dicFormFactorGeneral} does not converge and it needs to be subtracted. If  $1 \le \alpha(0)<2$, one subtraction is enough to make the integral in~\cref{eq:dicFormFactorGeneral} converge. However, there are two distinctive cases. If $\alpha(0) = 1$ we find $\tilde{F}_R(s) \sim \ln \left( \ln s \right)$ while if $1 < \alpha(0)<2$ the result reads $\tilde{F}_R(s)\sim s^{\alpha(0)-1}/\ln s$. The result can be generalized to $\alpha(0)>2$ with an arbitrary number of subtractions.

\section{Reggeized model with softening}\label{sec:regsoft}
The evaluation of \cref{eq:softened_regge} can be done using the relation
\begin{align}
\left[ G(\tilde p_2,m_q) \right]^n = \lim_{\bar m_q^2 \to m_q^2} \, \frac{(-1)^{n-1}}{(n-1)!} \left[\frac{d }{d \bar m_q^2 } \right]^{n-1} G(\tilde p_2,\bar m_q),
\end{align}
so the discontinuity can be written in terms of the \mbox{$(n-1)$th derivative}
\begin{align}
\Delta F_{R,n}(s)& =  \frac{(-1)^n }{16\pi\, s\, (n-1)!} \nonumber \\
\times  &\left[\frac{\partial }{\partial \bar m_q^2 } \right]^{n-1}_{\bar m_q^2 = m_q^2} \left[\lambda^{1/2}\left(s,m_q^2,\bar m_q^2\right)\int_{-1}^1 dz \, s^{\alpha( t )} \right]  , 
\label{eq:discreggesoft}
\end{align}   
where 
\begin{align}
t =  \frac{2 m_\pi^2 -  m_q^2 - \bar m_q^2  - s + z \rho_{2\pi} \lambda^{1/2}(s,m_q^2,\bar m_q^2)}{2}.
\end{align}

The discontinuity is fed into~\cref{eq:dicFormFactorGeneral} to provide the form factor.

\bibliographystyle{apsrev4-1}
\bibliography{refs}
\end{document}